\documentclass{aa}  

\usepackage{graphicx}
\usepackage{txfonts}
\usepackage{subfig}

\usepackage{color, colortbl}
\usepackage{hyperref}
\usepackage{xcolor}
\hypersetup{
  colorlinks,
  citecolor=darkred,
  linkcolor=red,
  urlcolor=blue}
\usepackage{amstext}

\begin{document} 

%
\definecolor{bubbles}{rgb}{0.91, 1.0, 1.0}
\definecolor{columbiablue}{rgb}{0.61, 0.87, 1.0}
\definecolor{cream}{rgb}{1.0, 0.99, 0.82}
\definecolor{lightblue}{rgb}{0.68, 0.85, 0.9}
\definecolor{lightcyan}{rgb}{0.88, 1.0, 1.0}
\definecolor{airforceblue}{rgb}{0.36, 0.54, 0.66}
\definecolor{armygreen}{rgb}{0.29, 0.33, 0.13}
\definecolor{bazaar}{rgb}{0.6, 0.47, 0.48}
\definecolor{cadetblue}{rgb}{0.37, 0.62, 0.63}
\definecolor{cadmiumgreen}{rgb}{0.0, 0.42, 0.24}
\definecolor{calpolypomonagreen}{rgb}{0.12, 0.3, 0.17}
\definecolor{celestialblue}{rgb}{0.29, 0.59, 0.82}
\definecolor{ceruleanblue}{rgb}{0.16, 0.32, 0.75}
\definecolor{cobalt}{rgb}{0.0, 0.28, 0.67}
\definecolor{coolblack}{rgb}{0.0, 0.18, 0.39}
\definecolor{darkpowderblue}{rgb}{0.0, 0.2, 0.6}
\definecolor{darkred}{rgb}{0.55, 0.0, 0.0}
\definecolor{deepcarmine}{rgb}{0.66, 0.13, 0.24}
\definecolor{ferngreen}{rgb}{0.31, 0.47, 0.26}

   \title{Near infrared and optical emission of WASP-5~b}

   \author{G. Kovacs\inst{1}
          \and
	  I. D\'ek\'any\inst{2}
	  \and
          B. Karamiqucham\inst{1,8}
	  \and
	  G. Chen\inst{3}
	  \and
	  G. Zhou\inst{4,5}
	  \and
	  M. Rabus\inst{6}
	  \and
	  T. Kov\'acs\inst{7}
          }

   \institute{Konkoly Observatory,  Research Center for Astronomy and Earth Sciences, 
              E\"otv\"os Lor\'and Research Network Budapest, Hungary
              \email{kovacs@konkoly.hu}
	      \and
	      Astronomisches Rechen-Institut, Zentrum f\"ur Astronomie der Universit\"at 
	      Heidelberg, Germany
              \and
	      Key Laboratory of Planetary Sciences, Purple Mountain Observatory, Chinese Academy of Sciences, Nanjing 210023, China
	      \and
	      Center for Astrophysics, Harvard \& Smithsonian, Cambridge, MA, USA
	      \and
	      Centre for Astrophysics, University of Southern Queensland, Toowoomba, Australia
              \and
	      Departamento de Matem\'atica y F\'isica Aplicadas, Facultad de Ingenier\'ia, 
	      Universidad Cat\'olica de la Sant\'isima Concepci\'on, Concepci\'on, Chile
              \and
	      Institute of Physics, Faculty of Science, E\"otv\"os Lor\'and University, 
	      Budapest, Hungary
	      \and
	      School of Physics, University of New South Wales, Sydney, Australia
             }

   \date{Received January 17, 2022; accepted ??, 2022}


%
%
  \abstract
{Thermal emission from extrasolar planets makes it possible to study 
important physical processes in their atmospheres and derive more 
precise orbital elements.}
{By using new near infrared and optical data, we examine how these 
data constrain the orbital eccentricity and the thermal properties 
of the planet atmosphere.}
{The full light curves acquired by the TESS satellite from two 
sectors are used to put upper limit on the amplitude of the 
planet's phase variation and estimate the occultation depth. 
Two, already published and one, yet unpublished followup observations 
in the 2MASS K (Ks) band are employed to derive a more precise 
occultation light curve in this near infrared waveband.} 
{The merged occultation light curve in the Ks band comprises 
$4515$ data points. The data confirm the results of the earlier 
eccentricity estimates, suggesting circular orbit: $e=0.005\pm 0.015$. 
The high value of the flux depression of $(2.70\pm 0.14)$~ppt 
in the Ks band excludes simple black body emission at the $10\sigma$ 
level and disagrees also with current atmospheric models at the 
$(4-7)\sigma$ level. 
From the analysis of the TESS data, in the visual band we found 
tentative evidence for a near noise level detection of the secondary 
eclipse, and placed constraints on the associated amplitude of the 
planet's phase variation. A formal box fit yields an occultation 
depth of $(0.157\pm 0.056)$~ppt. This implies a relatively high  
geometric albedo of $A_g=0.43\pm 0.15$  for fully efficient 
atmospheric circulation and $A_g=0.29\pm 0.15$ for no circulation 
at all. No preference can be seen either for the oxygen-enhanced, 
or for the carbon-enhanced atmosphere models. 
}
   {}


\titlerunning{Near infrared and optical emission of WASP-5~b}
\authorrunning{Kovacs et al.}

   \maketitle
%
%
%
%
\section{Introduction}
\label{sect:intro}
The year 2005 marks the first direct detection of the light radiated 
by an extrasolar planet \citep{charbon2005, deming2005}. The observations 
were made by the Spitzer space telescope \citep{werner2004} at $4.5$, $8$ 
and $24$~$\mu m$, not easily accessible by ground-based instruments. 
Although it was quite expectable that a similar measurement in the near 
infrared could also be possible by ground-based $4~m$-class telescopes, 
two years passed until the first tentative observation of that kind 
\citep{snellen2007}. Since then, secondary eclipse (occultation) 
observations at the $2.2$~$\mu m$ (2MASS K -- or Ks) band still remained 
in the realm of ground-based instruments, due to the lack of space 
instruments at this wavelength \citep[e.g.,][]{croll2015,zhou2015,martioli2018}. 
The 2MASS bands are especially suitable for the observation of hot 
extrasolar planets, due to the expected peaking of the black-body flux 
in $\sim1-2$~$\mu m$ for temperatures between $1500$ and $2000$~K, i.e., 
for the characteristic equilibrium temperatures of extrasolar planets 
\citep[e.g., ][]{alonso2018}. 

Here we revisit WASP-5, an ``ordinary'' extrasolar planetary system, 
discovered by the SuperWASP collaboration \citep{anderson2008}. The 
system harbors a single planet with a main sequence host, akin to our 
Sun. So far, no other planets have been reported in the system, although 
there are contradictory results concerning the origin of the transit 
time variation of planet {\em b} \citep[i.e., ][]{fukui2011, hoyer2012}. 
Based on the followup work of \cite{gillon2009}, the main system 
parameters are as follows: 
$R_s/R_\odot=1.029$, $M_s/M_\odot=0.960$, $T_{eff}=5700$~K, 
$a=0.0267$~AU, $R_p/R_J=1.087$, $M_p/M_J=1.58$. 
These parameters imply an equilibrium temperature (assuming 
zero albedo and full heat redistribution) of $1740$~K \citep{chen2014}. 
The orbital period is $1.6284300$~d, derived in this paper from the 
combination of the earlier epochs and those resulting from the 
analysis of the data from the TESS satellite \citep{ricker2015}. 

Occultation observations in the Ks band have already been carried out 
by \cite{chen2014} and \cite{zhou2015}. Here, we combine these data with 
our unpublished observations made by the $6.5$-m Walter Baade Telescope 
at the Las Campanas Observatory.\footnote{The Ks band photometric time 
series used in this paper are available at the CDS via anonymous ftp 
to cdsarc.u-strasbg.fr (130.79.128.5) or via 
http://cdsarc.u-strasbg.fr/viz-bin/qcat?J/A+A/3digitVol/Apagenumber}

Our main goal is to increase the precision of the estimation of 
the occultation depth -- an important ingredient for a more reliable 
model fitting, as most of the observations (including ours) sample 
the planetary spectra only in few, isolated bands. To constrain the 
atmosphere further, we utilize the recent data collected by the 
TESS satellite. We search for reflected light variation and 
occultation event. As a by-product of our analysis, we search also 
for additional planets (and find none).

%
%
\section{Datasets}
\label{sect:data}
Two occultation light curves in the near infrared Ks-band have been 
published so far on WASP-5~b. \cite{chen2014} used the MPG/ESO $2.2$~m 
telescope to observe the target in all three 2MASS bands. In spite 
of the substantial instrumental systematics they clearly detected 
the event after applying corrections due to positional and image 
quality dependences. \cite{zhou2015} performed a survey of seven 
hot Jupiters by using the Anglo-Australian Telescope (AAT). Their 
survey included also WASP-5, yielding a long-stretched coverage, 
allowing sufficient baseline in the eclipse modeling. Details of 
the observational settings and the methods used are described in 
the corresponding papers. 

%
%
\begin{table*}[t!]
\centering
\caption{Journal of K-band observations on WASP-5}
\label{obs}
\scalebox{1.0}{
\begin{tabular}{cccrllll}
\hline
Set & Date [UT]  & Dur. [h] & N    & Exp. [s] & Observer  & Ref. & Instr. / Telescope / Site\\
\hline 
1 & 08-09-2011 & 4.60         & 699  & $12.0$                 & Chen$^\bullet$       & \cite{chen2014} & GROND / MPG / ESO 2.2 / La Silla (Chile)\\
2 & 09-11-2011 & 4.33         & 2084 & $\phantom{1}4.4$       & D\'ek\'any$^\star$ & this paper & FourStar / Baade 6.5 / Las Campanas (Chile)\\
3 & 14-09-2014 & 5.99         & 1732 & $10.0^\blacktriangle$  & Zhou           & \cite{zhou2015} & IRIS2 / AAT 3.9 / Siding Spring (Australia)\\
\hline
\end{tabular}}
\begin{flushleft}
\vspace{-5pt}
{\bf Notes:}
All data were taken in the 2MASS K color, except for set-1, where the 
custom made K filter of the GROND instrument was used (with a 
transmission curve very close to that of the 2MASS K band).      
$^{\bullet}$Technical assistance is provided by Timo Anguita 
\citep[see ][]{chen2014}. 
$^\star$Assisted by Markus Rabus. 
$^\blacktriangle$Typical integration time. 
\end{flushleft}
\end{table*}

The third dataset comes from our single-night observations on 
9 November, 2011 (UT). The four-chip camera of the FourStar infrared 
imager attached to the $6.5$-m Walter Baade telescope was used to 
gather high cadence Ks images on the $10.9'\times10.9'$ field hosting 
WASP-5. An integration time of $4.4$~s was used, yielding a $\sim 7$~s 
overall sampling interval. For better photometric accuracy, the 
telescope was slightly defocused, resulting in stellar images of 
$\sim 10\arcsec$ diameter. All images were taken in a simple staring mode, 
without dithering. Unfortunately, the sky was not photometric 
throughout the night, due to intermittent clouds. This has led to  
losing some $300$ data points primarily after the ingress, affecting 
$\sim 25$\% of the full observing run.

%
%
\begin{figure}[h]
\centering
    \subfloat{{\includegraphics[width=0.20\textwidth]{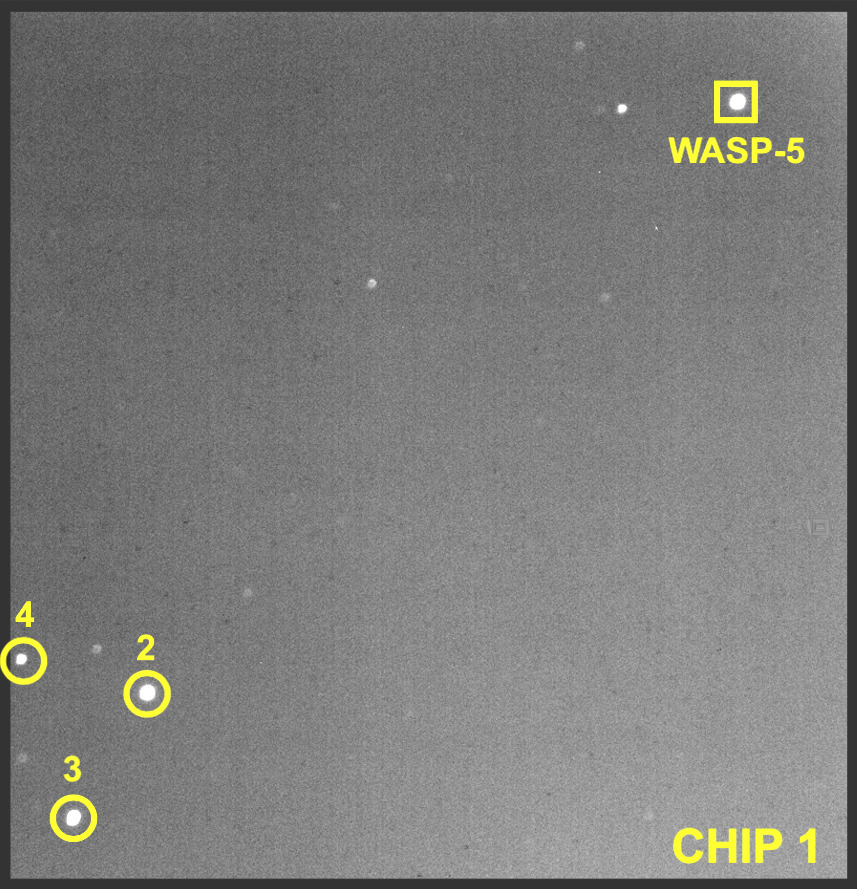}}}
    \quad
    \subfloat{{\includegraphics[width=0.25\textwidth]{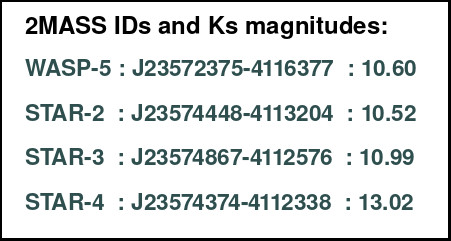}}}
\caption{The field of the first chip of the FourStar infrared mosaic imager of 
         the Baade telescope. North is to the left, East is to the bottom. 
	 Chips 2, 3 and 4 are located clockwise starting in the bottom of chip 1. 
	 The image size is $5.5'\times 5.5'$. We used comparison stars No. 2 
	 and 3 only.} 
\label{chip1_image}
\end{figure}

To obtain the photometric fluxes, we employed both the classical 
{\sc iraf}\footnote{{\sc iraf} is distributed by the National Optical 
Astronomy Observatories, which are operated by the Association of 
Universities for Research in Astronomy, Inc., under cooperative 
agreement with the National Science Foundation.} routines and those 
of the {\sc fitsh}\footnote{\url{https://fitsh.net/}} package by 
\cite{pal2012}. The two methods have led to very similar results, 
so we decided to use our earlier reduction made by {\sc iraf}. 

First we performed the standard reduction steps of bias, dark and 
flat corrections, including a treatment for the overall infrared 
sky emissivity variation by a nonlinear iterative multistep method,  
the nebulosity filtering 
algorithm\footnote{\url{http://casu.ast.cam.ac.uk/publications/nebulosity-filter/nebulosity_filter.pdf}} 
of \cite{irwin2010}. 
Then, we tested several aperture sizes to select 
the one that yielded the least scatter in the corresponding ensemble 
light curves (LCs). It turned out that nearly all apertures yield 
the same quality LCs, with a slight preference toward the mid-sized 
apertures. Finally, we selected the one with the aperture radius of 
$30$ pixels ($4.8\arcsec$), outer annulus starting at pixel radius of 
$40$ and ending at $50$, to assess the temporal background level.

In deriving the final ensemble LC (i.e., the target flux divided by 
the simple sum of the fluxes of the comparison stars), we decided not 
to use any comparison star from chips others than chip~1, that hosts 
the target. On this chip, (see Fig.~\ref{chip1_image}) we have two 
bright comparison stars (No. 2 and 3) and a fainter one (No. 4). 
We found that adding the fainter star slightly increases the 
noise\footnote{This is because the noise is not Poissonian. For faint 
objects, important contribution comes from the atmosphere, that acts 
on the derived fluxes of faint stars more violently, due to the increased 
significance of the background noise.}, therefore, we settled with 
the ensemble of the two brightest stars only.

%
%
\section{Merging the three Ks light curves}
\label{sect:3LC}
Before some of the peculiarities of the merging process are detailed, 
we describe the steps leading to the ensemble LC of the FourStar/Baade 
data (set-2 in Table~\ref{obs}). 

%
%
\subsection{The FourStar/Baade light curve}
\label{sect:set-2}
\begin{figure}[h]
\centering
\includegraphics[width=0.40\textwidth]{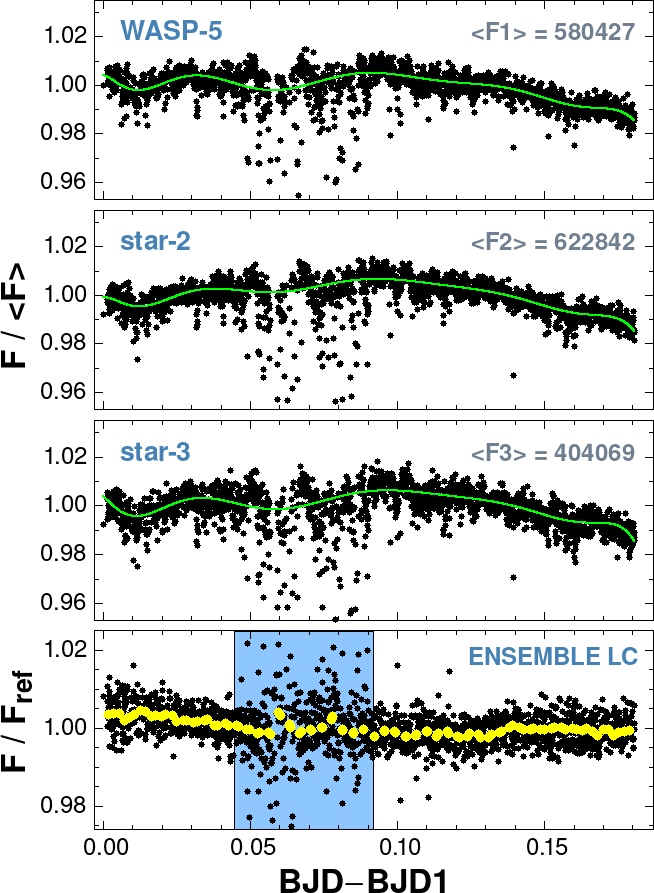}
\caption{Raw flux variations for set-2 (see Table~\ref{obs}) and 
         the resulting ensemble light curve (target flux 
	 over reference flux, $\sim F1/(F2+F3)$, without outlier 
	 correction, normalized to its average). Star-2 and 3 
	 (see Fig.~\ref{chip1_image}) served as comparison stars. 
	 The 11-th order polynomials, robustly fitted to the 
	 fluxes to handle outliers, are shown by green lines. 
	 Shaded area in the bottom panel indicates the period of 
	 intermittent clouds. The binned light curve (with 
	 overlapping bins -- see text) is shown by yellow dots. 
	 The time axis is shifted to the moment of the first data 
	 point (BJD1).}
\label{flux3}
\end{figure}

As mentioned, the biggest issue with the data is the temporal 
cloudiness during some part of the first half of the observation. 
The top three panels of Fig.~\ref{flux3} show the flux variation 
for the entire run, including the target and the two comparison 
stars. It is worth noting that for the better visibility of the 
part of the flux variation that is dominated by the non-outlying 
points, we limited the plots at the $4$\% flux drop. Several 
data points reach as much as $60$--$80$\% drops. 

%
%

Although the comparison stars serve an excellent diagnosis on the 
environmental origin of the harsh variations seen in the target, 
and the ensemble flux ratio cures most of the variation, we see 
in the bottom panel that the large drops in the flux could not be 
filtered out at the level required by the small signal we are 
searching for. Nevertheless, the binned LC strongly suggests the 
presence of an underlying occultation signal. We note that in 
constructing the binned LC, we used overlapping bin sets with 
a shift of half of the bin width. In this way we can test the 
dependence of the binned LC on the bin distribution, which is 
an important piece of information on the sensitivity of any 
conclusion to be drawn from the binned LC -- even if the 
conclusion is only preliminary. 

In processing further the set-2 data, we observe the following:  
1) there are outlier data points that are concentrated in a 
sufficiently broad section of the full time series and, therefore, 
might seriously bias the derived eclipse parameters; 
2) likely because of (however small) differential extinction, 
there is a significant downward trend in the ensemble LC. 
This should also be filtered out. 
3) A closer inspection of the ensemble LC in the bottom panel 
of Fig.~\ref{flux3} reveals that roughly in the middle of the 
cloudy period the flux suddenly jumped by a small fraction, 
enough to make a visible effect on the expected shallow eclipse. 
The most likely cause of this jump is the short-time change in 
the telescope pointing, leading to a sudden variation in the 
ensemble of pixels used in the flux evaluation. Leaving this 
jump in the ensemble LC would bias the occultation depth. 

%
%
\begin{figure}[h]
\centering
\includegraphics[width=0.40\textwidth]{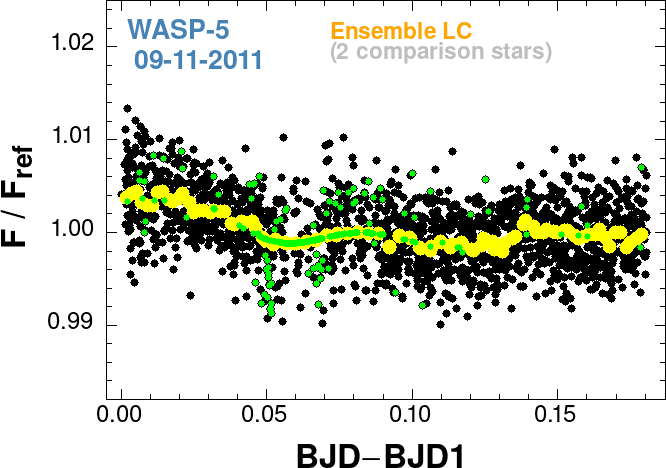}
\caption{Ensemble LC obtained from set-2 after polynomial outlier 
         correction. Black dots are original data points. 
	 Green dots are from the polynomial corrections. Yellow 
	 dots are from binning the data points independently 
	 of their origins (corrected or not corrected).}
\label{trimmed_LC}
\end{figure}

By following the principle of `least data massaging', we 
proceeded as follows. For the treatment of outliers (issue 1)  
we robustly fitted\footnote{In the standard least squares 
fit we employed Cauchy weights, adjusted iteratively to the 
processed time series -- see, e.g., \cite{kovacs2020}.} 
$11^{\rm th}$-order polynomials to the fluxes of the target 
and the comparison stars. After the fit we employed a 
$3\sigma$-clipping for the outliers and replaced these items 
by the corresponding polynomial values for the respective fluxes. 
In this way we naturally ended up with an ensemble LC that had 
no outliers, however, showed some trace of the `trimming' made. 
Figure~\ref{trimmed_LC} displays where the polynomial replacement 
of the original data points were made (green dots). When both 
the target and all of the comparison stars had to be corrected, 
we see a continuous sequence of points. In all other cases the 
corrected points scatter around the ridge, represented by the 
binned LC (yellow points).  

The linear trend and the jump in the ensemble LC (issues 2 and 3)  
were treated within an iterative process by filtering out these 
systematics, fitting the cleaned LC to an eclipse model and then 
subtracting this eclipse model from the starting dataset to get 
the next approximation for the systematics. 

The systematics were represented by a linear function for the 
trend and a jump function to handle the discontinuity 
mentioned above:
%
%
\begin{eqnarray}
\label{trend}
F(t) = c_0 + c_1t + c_2 H(t_{jump}) \hspace{2mm},  
\end{eqnarray}
where $F$ is the observed flux, $t$ is the time, measured from 
the first data point. $H$ is the Heaviside function with unit step 
at $t_{jump}=0.068$~d. The jump position was fixed throughout 
the fit. Because the star blocks all radiation from the planet, 
the trapezoidal approximation for the occultation light curve 
suits perfectly: 
%
%
\begin{eqnarray}
\label{eclipse}
T(t) = 
\begin{cases}
  1 & \text{if $t\leq t_1$ or $t\geq t_4$} \\
  1-\delta & \text{if $t_2 \leq t \leq t_3$} \\
  1-\delta \times (t-t_1)/\Delta t & \text{if $t_1 \leq t \leq t_2$} \\
  1-\delta \times (t_4-t)/\Delta t & \text{if $t_3 \leq t \leq t_4$}
\end{cases}
\hspace{2mm},  
\end{eqnarray}
where $t_1$, $t_2$, $t_3$ and $t_4$, respectively, are the moment 
of ingress (first contact), start and end of the total eclipse and 
the moment of egress. The length of the ingress and egress phases 
are assumed to be equal: $t_2-t_1 = t_4-t_3 = \Delta t$. Except 
for the eclipse depth $\delta$, all these parameters are scanned 
for the best fit within the framework of robust least squares. For 
any given set of $\{t_i\}$ the transit depth was fitted in one step, 
due to the linear nature of the parameter. The systematics parameters 
$\{c_i\}$ were fitted in the same manner. The final light curve for 
set-2 are shown in Sect.~\ref{sect:set-123}

%
\subsection{The three light curves}
\label{sect:set-123}
In trying to treat all three datasets in the same way, i.e., by 
starting from the simple ensemble light curve and employing the 
``minimum massage'' post processing step, we found that the case 
of set-1 \citep{chen2014} is different. Since the ensemble light 
curve suffers excessively from systematics (see Fig.~2 of that paper), 
we decided to use their processed light curve that was obtained 
by applying carefully chosen external parameters (such as stellar 
position and image size) to separate systematics. On the other 
hand, for set-3 of \cite{zhou2015} we used their simple ensemble 
light curve, even though there is also a substantial nonlinear 
trend in the data. In spite of this, we decided not to use an 
airmass or some polynomial correction (as given in the original 
paper), since this may introduce unpredictable changes in the 
eclipse and significantly depress the depth of the occultation. 

\begin{figure}[h]
\centering
\includegraphics[width=0.40\textwidth]{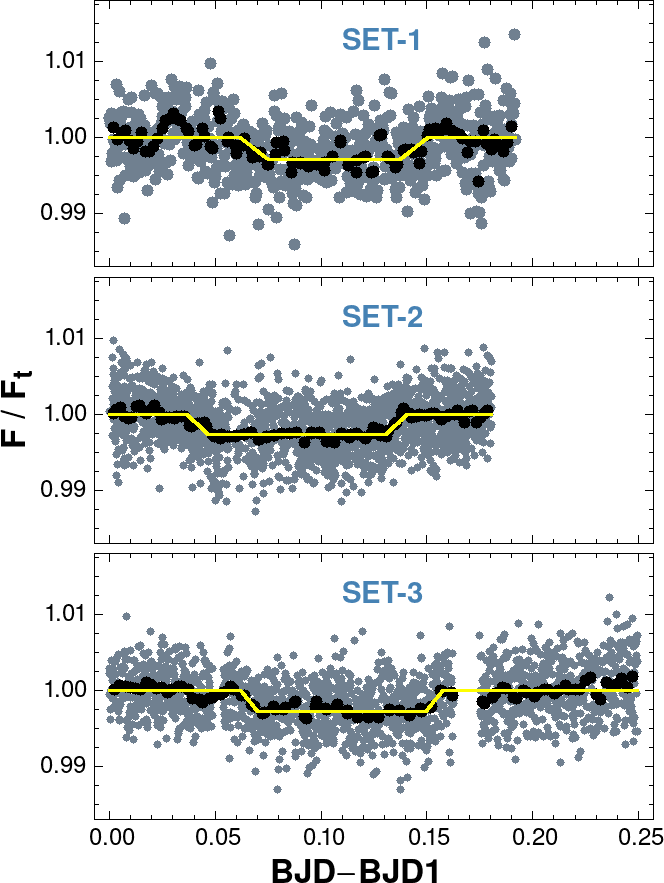}
\caption{The three light curves of Table~\ref{obs} in the 
         pre-merging phase. All light curves are filtered out from 
	 time-dependent linear trends and normalized by the 
	 total (star$+$planet) flux F$_{\rm t}$. Black dots 
	 are the binned values, yellow lines are the trapezoidal 
	 models fitted to the original (unbinned) data shown by 
	 deep gray dots. For better visibility we increased the 
	 point size for set-1.}
\label{LC-123}
\end{figure}

All three light curves (sets-1 and -3 as above, set-2 as given 
in Fig.~\ref{trimmed_LC}) serve as the input time series to fit 
them individually by the eclipse model and a linear trend 
(extended by a jump function for set-2). The result of this 
procedure is shown in Fig.~\ref{LC-123}. 

%
%
%

%
\subsection{Merging the three light curves}
\label{sect:merge-123}
Before constructing the merged light curve, we need to check 
if such a merging is possible, i.e., if there is a unique 
orbital period that matches all three light curves within 
the observational errors. Evaluation of the updated orbital 
period by using the primary transit observations from the 
TESS satellite and combining the ephemerides with earlier 
followup data will be given in Sect.~\ref{sect:tess}. Here 
we merely use the orbital period and the moment of the 
transit derived from that analysis. 

%
%
\begin{figure}[h]
\centering
\includegraphics[width=0.40\textwidth]{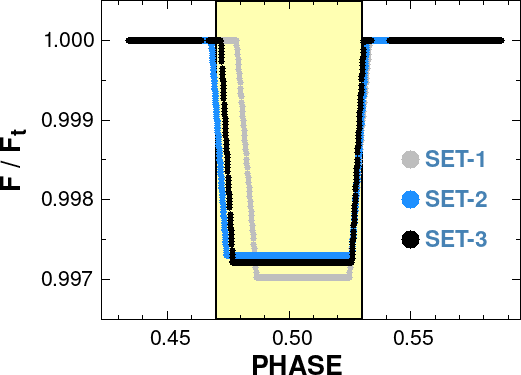}
\caption{Trapezoidal models fitted to the Ks observations and 
         folded by the orbital period. The shaded rectangle shows 
	 the expected event of occultation, assuming circular orbit 
	 and using the updated orbital period and transit center: 
	 $P=1.6284300$~d, $T_{cen}=2458355.50805$ [BJD] (see 
	 Sect.~\ref{sect:tess}).}
\label{t14-123}
\end{figure}
%

%
%
\begin{figure}[h]
\centering
\includegraphics[width=0.40\textwidth]{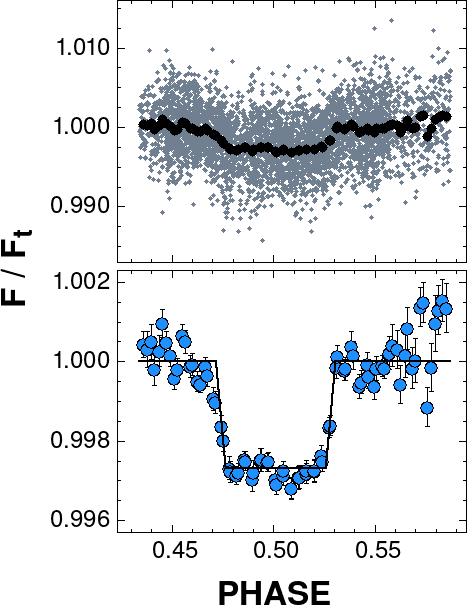}
\caption{{\em Upper panel:} Phase-folded occultation curve of all 
         the Ks observations with bin averages (black dots). 
	 {\em Lower panel:} Binned light curve of the above 
	 dataset with the errors of the bin averages and the 
	 trapezoidal model (black line) fitted to the original 
	 (unbinned) data shown in the upper panel.}
\label{merge_k}
\end{figure}

By fitting the individual folded light curves we can examine 
if the data suggest strong discrepancies signaling warnings 
to be considered during the merging process. Figure~\ref{t14-123} 
shows the individual phase-folded fits, indicating that the 
three datasets are in reasonable agreement, even if we consider 
set-1, the most discrepant from all. Set-1 contains the least 
number of data points ($699$ vs $2084$ and $1732$ for set-2 and 
3, respectively), and has also the largest residual (data minus 
fit) scatter (in relative flux units: $\sigma=0.0036$, vs $0.0030$ 
and $0.0033$). In spite of these differences, all three datasets 
yield remarkably close egress phases. The cause of this is not 
entirely clear at this moment. In some cases, it might be simply 
the sign of more stable sky conditions in the second part of 
the run (i.e., for set-2 this was indeed the case).    

In the final step of the merging process, we packed all data 
points in a single phase-folded dataset, by discarding the 
relatively small differences in data quality (i.e., weighting 
all data points from all sets equally).\footnote{This choice 
is partially justified, because of the compensating effect of the 
larger number of points for the datasets with somewhat lower noise.} 
The phase-folded light curve, containing all the $4515$ data points,   
was robustly fitted by the trapezoidal model. The resulting binned 
light curve and the best-fit trapezoidal are shown in Fig.~\ref{merge_k}. 
The fitted parameters are listed in Table~\ref{merged_par}. The 
errors were computed from simple Monte Carlo simulations, whereby 
the binned light curve (mapped back to all the $4515$ phase points) 
was perturbed by a bin-dependent Gaussian noise. We opted to 
use the binned time series rather than the trapezoidal fit, because 
of the remaining systematics, especially before/after the ingress/egress. 
We generated $500$ mock time series, fitted trapezoidals to each 
realization, and, after completion, we computed the standard 
deviations of the parameters. We refer to these standard deviations 
as the $1\sigma$ errors of the respective parameters. In Appendix 
\ref{app_A} we give further details of the error calculation and 
the improvement of the parameters by using the merged data as compared 
with the fits to the individual datasets. 

%
%
%
\begin{table}[h!]
\centering
\caption{Trapezoidal occultation parameters for all Ks data}
\label{merged_par}
\scalebox{1.0}{
\begin{tabular}{lcc}
\hline
Parameter  &  Value  &  Error\\
\hline
\hline
$T_{1}$    & 0.47146 & 0.00219\\
$T_{4}$    & 0.53102 & 0.00139\\
$T_{14}$   & 0.05956 & 0.00270\\
$T_{12}$   & 0.00528 & 0.00070\\
$\delta$   & 0.00270 & 0.00014\\
\hline
\end{tabular}}
\begin{flushleft}
\vspace{-5pt}
{\bf Notes:}
All eclipse times are in the units of the orbital phase. 
Eclipse depth $\delta$ is the relative flux depression. The 
ingress and egress times, $T_{1}$ and $T_{4}$ can be converted 
into Barycentric Julian Date (TDB standard) by using the 
following formulae, e.g., for the ingress: 
$T_{ing}$[BJD]$=T_{cen}+P\times(n+T_{1})$, where $n$ is the epoch 
number of the event of interest and $T_{cen}=2458355.50805$ 
is the moment of the transit center and $P=1.6284300$~d is 
the orbital period. Epochs are `as observed', i.e., no correction 
was made due to orbital light time effect of $27$~s. 
\end{flushleft}
\end{table}
%

%
\section{Analysis of the TESS data}
\label{sect:tess}
We use the light curves acquired by the full sky survey satellite 
TESS for: 
a) updating the ephemeris of the transit (since the occultation 
and the transit data were acquired in different epochs, we need a 
precise ephemeris to predict the transit phase right before 
the occultation occurred, if we want to make an estimation 
on the eccentricity);  
b) to measure the emission in the optical, we need space-based 
data, because of the high precision needed to detect the signature 
of the planet at this wavelength in the phase of occultation 
-- the thermal emission in the optical is small, due to the planet's 
low temperature and, in general, the albedos of the gas giants 
are also small \citep[e.g.,][]{wong2020, wong2021}. 

WASP-5 was observed by TESS in sector $02$ between August 22 and 
September 20, 2018. Then, the object was revisited while scanning 
sector $29$ between August 26 and September 22, 2020. The two 
segments comprise altogether over $30000$ data points in the 
short cadence ($2$~min) sampling rate. We note that \cite{wong2020} 
have already performed an analysis of the sector $02$ data and 
ended up with similar conclusions to ours as to be detailed 
in the subsections below.  

\begin{figure}[h]
\centering
\includegraphics[width=0.40\textwidth]{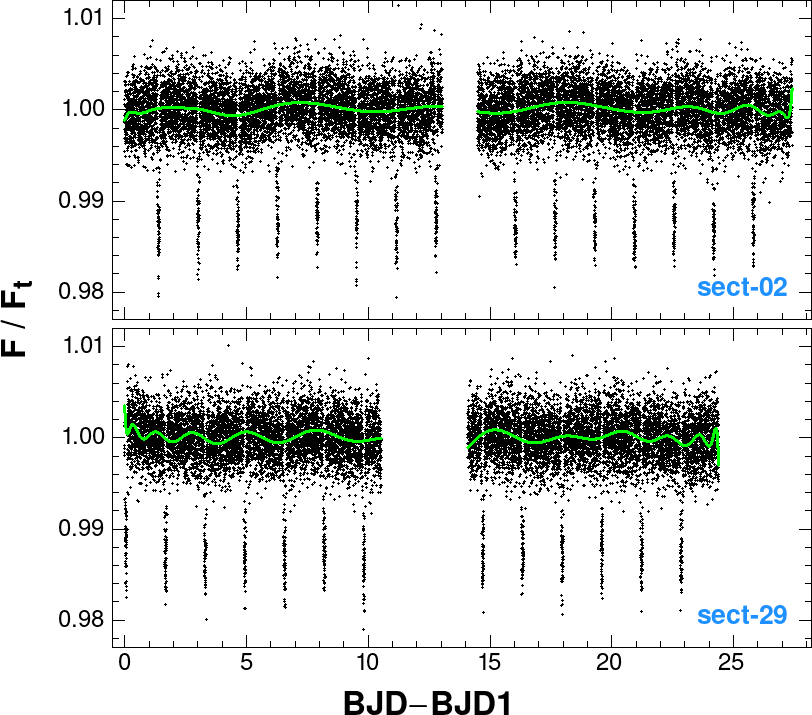}
\caption{TESS light curves of WASP-5 from the sectors shown in the 
         bottom right corners. The light curves have been processed 
	 by the TESS pipeline by using PDC systematics corrections 
	 on the SAP fluxes. The $36^{th}$-order polynomial fit is 
	 shown by green line, and used to filter out the remaining 
	 systematics and possible stellar variability.}
\label{tess_LC}
\end{figure}

Figure~\ref{tess_LC} shows the light curves from the above two 
sectors after employing the Presearch Data Conditioning (PDC) 
method of \cite{smith2012} and \cite{stumpe2012} implemented in 
the TESS pipeline\footnote{See the corresponding TESS manual
\url{https://heasarc.gsfc.nasa.gov/docs/tess/docs/jenkinsSPIE2016-copyright.pdf}}. 
The Simple Aperture Photometric (SAP) time series served as the 
input for the PDC filter. Both types of data were downloaded from 
the STScI MAST site.\footnote{\url{https://archive.stsci.edu/hlsp/search.php}} 
We filtered the data further by using a $36^{th}$-order robust 
polynomial fit to minimize the effect of the remaining systematics 
and possible stellar variability. The effect of this filtering is 
discussed in Appendix~\ref{app_B}. 

Because sector $02$ SAP data suffer from a large number of outlying 
data points, to make the analysis uniform, we performed an 
iterative $3\sigma$ clipping for all datasets. The clipping was 
made relative to the transit model and the clipped values were 
set equal to the corresponding model values.    

%
%
%

%
\subsection{Updating transit ephemeris}
\label{sect:tess-eph}
To derive transit light curves free from other variations, 
we employed the same type of robust iterative method as 
briefly described in Sect.~\ref{sect:set-2}. The input data 
were the PDC/SAP time series as mentioned above. The model 
time series constituted two multiplicative parts: the transit   
and a $36^{th}$-order polynomial. For the transit we adopted 
the simple model of \cite{kovacs2020}, representing the 
ingress/egress phases as linear flux depressions with the same 
steepness and duration. The limb darkening was modeled by 
a scalable U-shaped function. We found this model quite 
satisfactory at the level of the accuracy of the data analyzed.  

%
%
%
\begin{table}[h!]
\centering
\caption{Transit parameters of WASP-5 from two TESS sectors}
\label{tess_transits}
\scalebox{0.9}{
\begin{tabular}{cccccc}
\hline
Sect.  &  Type & $T_{cen}$ [$\rm BJD_{TDB}$] & $T_{14}$ [d] & $T_{12}$ [d] & $\delta$ [flux]\\
\hline
\hline
02 & SAP &  2458355.50802 & 0.09870 & 0.01091 & 0.01335 \\
02 & PDC &  2458355.50807 & 0.09855 & 0.01083 & 0.01340 \\
29 & SAP &  2459088.30168 & 0.09781 & 0.01082 & 0.01382 \\
29 & PDC &  2459088.30166 & 0.09673 & 0.01017 & 0.01380 \\
\hline
\end{tabular}}
\begin{flushleft}
\vspace{-5pt}
\end{flushleft}
\end{table}

The transit parameters for the various time series are shown in 
Table~\ref{tess_transits}. After combining these with the transit parameters 
obtained from the five followup observations of \cite{baluev2019}, 
\cite{moyano2017}, \cite{hoyer2012}, \cite{fukui2011} and \cite{anderson2008}, 
we found that the orbital period of \cite{fukui2011} should be decreased by 
$0.123$~s to properly match the published epochs.\footnote{When using the 
period of \cite{fukui2011} we get an overall difference of $\sim 4$~min, 
whereas with the $0.123$~s lower period the differences are below $1$~min 
and mostly $0.5$~min.} By choosing sector $02$ timing as a reference, the 
final ephemeris is given in Table~\ref{eph_update}. The error of the epoch 
was computed from $50$ simple Monte Carlo simulations by using the PDC 
data and is equal to the standard deviation of the epochs obtained from 
the $50$ realizations. The error on the period was calculated from 
$(\sigma_1^2+\sigma_2^2)^{1/2}/2444$, where $\sigma_1=0.00019$~d as given 
in \cite{fukui2011}, $\sigma_2$ is the epoch error as given in 
Table~\ref{eph_update} and the integer in the denominator is the elapsed 
epoch number between the two epochs. It is worth noting that the 
currently published ephemerides by \cite{ivshina2022} are in complete 
agreement with ours. There are $7$~ms and $17$~s differences between the 
periods and transit centers, respectively, corresponding an agreement 
within $1-2\sigma$. 

%
%
\begin{table}[h!]
\centering
\caption{Updated orbital period and mid-transit time for WASP-5}
\label{eph_update}
\scalebox{1.0}{
\begin{tabular}{rr}
\hline
$P_{orb} [d]$  & $T_{cen}$ [$\rm BJD_{TDB}$] \\
\hline
\hline
$    1.62843000$     & $2458355.50805$ \\
$\pm 0.00000009$     & $\pm   0.00013$ \\
\hline
\end{tabular}}
\begin{flushleft}
\vspace{-5pt}
{\bf Notes:}
$T_{cen}$ resulted from the analysis of the current (2018 and 2020) 
TESS visits, the period was derived from the combination of these 
TESS data and earlier followup observations dated back to the discovery 
\citep{anderson2008} of WASP-5.  
\end{flushleft}
\end{table}
%

%
\subsection{Search for occultation and phase variation}
\label{sect:tess-occ}
To test the dependence of a possible detection of these delicate 
features on the data processing methods, we used four data types: 
SAP light curve with or without robust polynomial correction 
(see Sect.~\ref{sect:tess-eph}); 
PDC light curve with the same options. 
After prewhitening by the transit, we performed a bin signal 
search in the light curve folded by the orbital period. To account 
for the possible other variations, we employed a fully binned 
analysis, where the out of eclipse region was also divided into 
bins of the same size as the eclipse duration. After the bin with 
the largest flux depression was identified, we used a simple 
statistic to characterize its significance. Similarly, the 
phase variation was studied simply by a single-component Fourier 
fit to the original (i.e., not binned) phase-folded light curve. 
Significance tests were performed by using injected signals into 
pure Gaussian time series. Further details on the secondary eclipse 
and phase variation searches together with the supplementary 
statistical tests are given in Appendix~\ref{app_B}. Here we 
summarize the constraints derived in that appendix. 

\begin{figure}[h]
\centering
\includegraphics[width=0.40\textwidth]{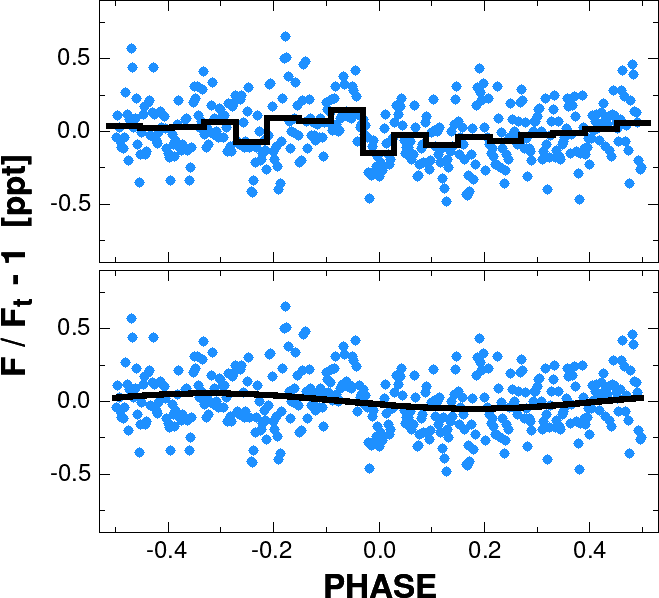}
\caption{Phase-folded and binned light curve of WASP-5 for the 
         full TESS dataset after dividing the PDC flux values by 
	 the transit and polynomial components. The upper and lower 
	 panels, respectively, show the bin and cosine models 
	 (black lines). The transit phase is shifted to $-0.5$ 
	 for better visibility of the neighborhood of the expected 
	 phase of the secondary eclipse.}
\label{occ_LC}
\end{figure}

First, for the illustration of the data quality at the expected 
level of reflected light variation, we show the transit- and 
polynomial-filtered, PDC-processed light curve in Fig.~\ref{occ_LC}. 
The blue dots resulted from overlapping binning 
(see Sect.~\ref{sect:set-2}) with $200$ bins ($400$ points altogether). 
The bin model has a bin width equal to the transit length, 
yielding 17 bins. Although this figure has little indication for 
the presence of the type of signals we are searching for, as shown 
in Appendix~\ref{app_B}, the parameters fitted to the data of 
various processing levels remain remarkably stable. This leads to 
the following average values of the secondary eclipse depth and 
phase variation amplitude: 
$\delta = 0.157\pm 0.056$~ppt, $2A = 0.113\pm 0.041$~ppt.

%
%

With additional statistical tests we found that at the observed 
amplitude of the cosine component, it has only $0.3$\% probability 
that the underlying phase variation has a total (peak-to-peak) 
amplitude greater than $0.20$~ppt. Also, for a boxy eclipse of 
the same depth, the probability that the bin model yields a phase 
solution outside the expected secondary eclipse phase is less 
than $10$\%. With the observed correct location of the main dip 
for all four datasets, this suggests that we may have found a 
signature of the underlying signal. 

Although the phase variation seems to yield a more stringent limit 
on the eclipse depth, the discordant phase\footnote{The cosine fit  
exhibits considerably lower phase stability than the bin fit even 
for simple white noise (see Appendix~\ref{app_B}). In addition,  
we may also have other sources (e.g., stellar variability, instrumental 
systematics) that interfere with the phase variation -- but, because 
of the different time scales, leave the secondary eclipse relatively 
intact.} of the cosine fit refrains us from relying too much on the 
result suggested by this fit. Therefore, we use the eclipse depth 
quoted above as our best guess at present for the real secondary 
eclipse depth in the TESS waveband.

%
\subsection{Search for additional transit components}
\label{sect:tess-multi}
Although hot Jupiters systematically avoid close planetary companions 
\citep{poon2021}, it is still a matter of interest if WASP-5 is one 
of those rare systems. Unfortunately, the short time spans of the 
TESS observations make the search for the more common longer period 
companions less trivial, leading to lower observed multiple system 
rates from the TESS data \citep[][]{otegi2021}. 

After prewhitening by the transit, we performed BLS searches 
\citep{kovacs2002} in the frequency interval $[0.01,10]$~c/d. 
The time series contains two dense tracks, separated by 
$\sim 730$~days, comprising $36778$ SAP data points altogether. 
We tested all four data type combinations (SAP, PDC with or without 
polynomial filtering). All data types show an increasing power 
excess from $1$~c/d down to $0.01$~c/d with no prominent peak in 
this frequency interval. The spectra are flat in $[1,10]$~c/d, 
without any dominant peak superposed on the white noise background. 

\begin{figure}[h]
\centering
\includegraphics[width=0.40\textwidth]{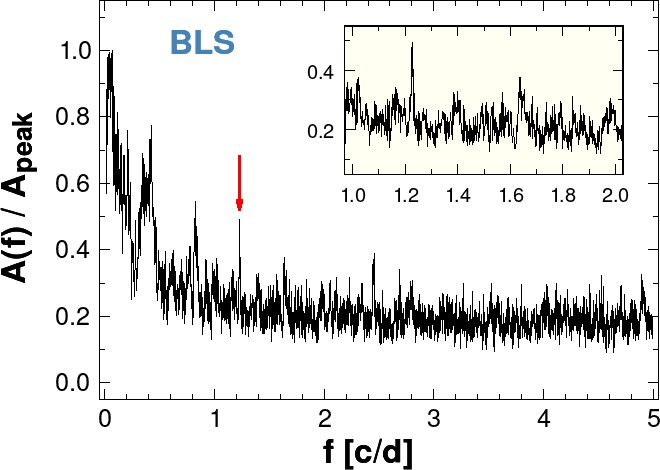}
\caption{Injected transit test of the full TESS dataset. We 
         used the SAP light curve to inject the signal and then 
	 robust polynomial filtering was employed to lower the 
	 the red noise. We show the frequency spectrum of the 
	 so-derived time series, after subtracting the transit 
	 signal of planet {\em b}. Red arrow indicates the peak 
	 due to the injected signal with a transit depth of 
	 $0.3$~ppt. The inset shows the close neighborhood of 
	 the test signal.}
\label{raw_bls_sp}
\end{figure}

To test the detection limit in the potentially interesting frequency 
interval of low-order resonance, we injected a transit signal in the 
original SAP time series. Then we performed a polynomial filtering 
as mentioned earlier in this section. The injected signal had a period 
of half of the orbital period of planet {\em b} and a transit depth 
of $0.3$~ppt (corresponding to $1.8$ Earth radii). We used a boxy  
transit with the same duration as that of planet {\em b}. The result 
is shown in Fig.~\ref{raw_bls_sp}. 

%
%

From the structure of the spectrum, it is clear that $0.3$~ppt 
transit depth is close to the low limit of a transit signal 
we can hope to detect in the available dataset. This limit is 
changing as a function of dataset and frequency, and is obviously 
higher for signals with periods longer than one day.

%
%
\section{The eccentricity}
\label{sect:ecc}
\begin{figure}[h]
\centering
\includegraphics[width=0.35\textwidth]{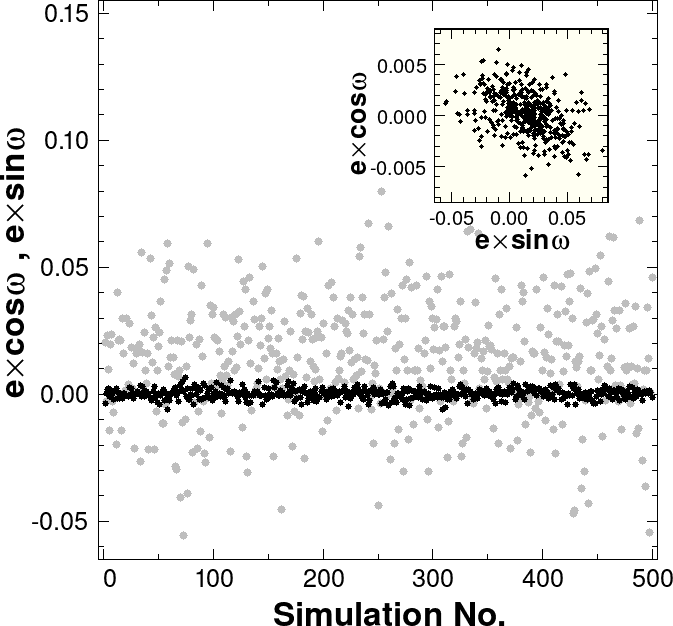}
\caption{Eccentricity components obtained from the Monte Carlo 
         simulations as described in the text. Gray and black 
	 dots, respectively, denote the $e\sin{\omega}$ and 
	 $e\cos{\omega}$ components. The inset shows the correlation 
	 between the two components, leading to a smaller error 
	 on the eccentricity as compared to that of the 
	 $e\sin{\omega}$ component alone.}
\label{ecos_esin}
\end{figure}

With the occultation ephemeris derived in Sect.~\ref{sect:3LC} and 
with the updated transit ephemeris by using the TESS data in 
Sect.~\ref{sect:tess}, we can easily compute the two components of 
the eccentricity. For an easier reference, the components are 
as follows \citep{winn2014} 
%
%
\begin{eqnarray}
\label{eq:ecc}
e\cos{\omega} & = & {\pi \over 2}(\varphi_{obs}-\varphi_{cal})  \hspace{2mm}, \\
e\sin{\omega} & = & {T_{14}(oc)-T_{14}(tr) \over T_{14}(oc)+T_{14}(tr)} \hspace{2mm},  
\end{eqnarray}
where $\varphi_{obs}$ and $\varphi_{cal}$, respectively, are the 
observed (corrected for light-time effect) and calculated  
phases of the occultation centers (the latter is with the 
assumption of circular orbit). The argument of periastron is denoted by 
$\omega$. As described in Sect.~\ref{sect:merge-123}, the errors of the 
occultation signal in the Ks band were computed from a simple Monte Carlo 
simulation, based on the binned version of the merged data from the three 
data sources. The observational noise was considered to be multiplicative 
and non-stationary, according to the standard deviations around the bin 
means. For the $500$ realizations we calculated the eccentricity components 
from the fitted trapezoidal occultation parameters. As it is obvious from 
Eq.~\ref{eq:ecc}, the two components are not independent. Furthermore, 
$e\cos{\omega}$ is expected to be less noisy, as several authors noted 
previously \citep[e.g., ][]{winn2014}. Indeed, Fig.~\ref{ecos_esin} 
clearly shows both the correlation and the considerably tighter behavior 
of $e\cos{\omega}$. We also observe that the $e\sin{\omega}$ component 
is shifted to more positive values. This is because noise makes the 
ingress/egress parts shallower, leading to the preference of longer 
eclipse durations in the best-fit search.    

%
%

From these simulations we obtained the errors also for the eccentricity 
components and, finally, for the eccentricity. For simple reference, we 
summarized these parameters in Table~\ref{tab:ecc}. The correlation 
between the two eccentricity components is also exhibited by the lower 
error on the eccentricity than that on the $e\sin{\omega}$ component. 

It is instructive to compare our eccentricity values with those derived 
from the Spitzer data by \cite{baskin2013}. First we checked if there 
was any difference between using our transit ephemerides vs those 
employed by \cite{baskin2013}. We found that for the two epochs 
\cite{baskin2013} published in their Table 1, our ephemerides predicted 
an average offset $1$~min greater than the one calculated from the 
ephemerides of \cite{fukui2011} ($3.7$~min vs $4.7$~min). Their 
offset time implies $e\cos{\omega}=0.0025\pm 0.0012$. Because 
the agreement is at the $\sim 1\sigma$ level between their value 
and ours, we can average them out and arrive at a value of 
$0.0020\pm 0.0016$, not implying anything different from 
zero eccentricity.\footnote{\cite{baskin2013} did not publish 
$e\sin{\omega}$ values, so we cannot compare the eccentricities 
directly.}

%
%
\begin{table}[h!]
\centering
\caption{Eccentricity from the occultation and transit parameters}
\label{tab:ecc}
\scalebox{1.0}{
\begin{tabular}{lrc}
\hline
Quantity  &  Value  &  Error \\
\hline
\hline
 $e\cos{\omega}$ & $ 0.00145$ & 0.00195 \\
 $e\sin{\omega}$ & $-0.00518$ & 0.02197 \\
 $e$             & $ 0.00538$ & 0.01520 \\
\hline
\end{tabular}}
\begin{flushleft}
\vspace{-5pt}
\end{flushleft}
\end{table}
%

%
\section{The emission spectrum}
\label{sect:em_spectr}
Here we examine how the more accurate occultation depth in the 
near infrared (Sect.~\ref{sect:merge-123}) and our preliminary 
estimate on the same quantity in the visible from the TESS data 
(Sect.~\ref{sect:tess-occ}) can constrain the atmospheric 
properties of WASP-5~b. The secondary eclipse analysis 
was presented in Sect.~\ref{sect:3LC}, where we derived an 
occultation depth of $\delta(occ,Ks)=(2.70\pm 0.14)$~ppt in 
the near infrared.

In the visible, corresponding to the wide-band filter 
of TESS\footnote{$\lambda_{eff}=0.746$, $W_{eff}=0.390$~$\mu$m, 
see:\\ \url{http://svo2.cab.inta-csic.es/theory/fps/}}, 
we use the average of the eclipse depths obtained from four 
types of datasets: $\delta(occ,vis)=(0.157\pm 0.056)$~ppt. Although 
we could use also the value obtained from the estimation of the 
phase variation, we opted not to use this value for reasons 
discussed in Sect.~\ref{sect:tess-occ}.   
 
\cite{baskin2013} have measured the planet's emission at 
$3.6$~$\mu$m and $4.5$~$\mu$m by the Spitzer infrared satellite. 
Although we do not make any model fitting in this paper -- since 
we use the same models as given by \cite{chen2014} --, we found 
it instructive to display all, currently available data on the 
same plot. 

\begin{figure}[h]
\centering
\includegraphics[width=0.40\textwidth]{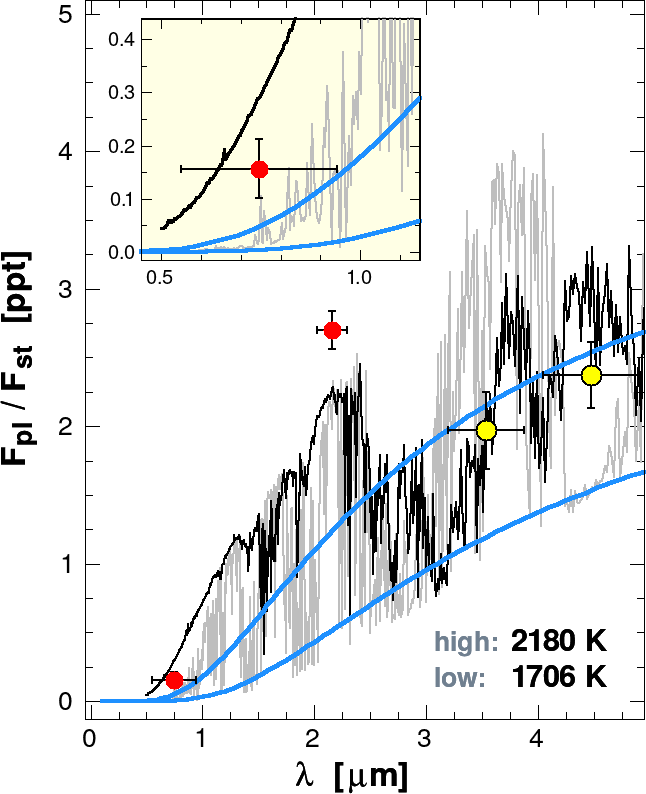}
\caption{Model thermal emission spectra from \cite{chen2014} 
         (gray for oxygen- and black for carbon-enhanced 
	 chemical compositions). Black body lines for fully 
	 efficient ($\alpha=0.25$) and non-efficient 
	 ($\alpha=2/3$) circulations are shown 
	 by blue lines. The corresponding temperatures 
	 are given in the lower right corner. The optical and 
	 infrared occultation depths derived in this paper 
	 are shown by red dots, the Spitzer data of 
	 \cite{baskin2013} by yellow dots. The inset zooms 
	 in the optical waveband.}
\label{em_spec}
\end{figure}

The atmospheric models presented by \cite{chen2014} are based on   
the plane-parallel equilibrium models of \cite{mandhu2009, mandhu2010}, 
employing free pressure--temperature profile and chemical composition. 
Figure~\ref{em_spec} shows these theoretical spectra and the black 
body lines for fully efficient and zero circulations  
\citep[i.e., lack of heat exchange between the day and night sides of the planet -- see][]{cowan2011}. 
The atmospheric models have monotonic pressure--temperature profile 
(i.e., no temperature inversion). The depth of the atmosphere was 
chosen to fit the brightness temperatures corresponding to the J, H, 
K data of \cite{chen2014}.      

There are two essential conclusions we can draw from the positions 
of the new data points in respect to these models. First, the lower 
error bar on Ks increased the significance of the higher Ks flux 
and suggests strong emission in this waveband. This can be 
realized by additional emitters at a deeper level of the atmosphere 
(corresponding to temperatures higher than $2700$~K). 

Second, even though our estimate for the secondary eclipse 
depth is only tentative in the TESS band, it still yields a useful  
piece of information. This is because of the relatively small error 
on the data with respect to the spectral features in the optical. 
In particular, the current value of the emission in the visible 
corroborates what the Spitzer data may also indicate, i.e., no strong 
preference for any of models used. 
%
%

From the optical occultation depth we can also estimate the geometric 
albedo. For an easier reference, here we repeat the necessary formulae  
presented by \cite{cowan2011} and, e.g., by \cite{daylan2021}. The 
observed occultation depth constitutes two parts: the thermal radiation 
by the planet and the reflected light of the host star
%
%
\begin{eqnarray}
\label{eq:tot_depth}
\delta_{obs} = \delta_{therm} + \delta_{refl} 
\hspace{2mm}.  
\end{eqnarray}
Assuming circular orbit, the reflected light is directly related 
to the geometric albedo $A_g$ 
%
%
\begin{eqnarray}
\label{eq:galbedo}
\delta_{refl} = A_g \left({R_p \over a}\right)^2 
\hspace{2mm},   
\end{eqnarray}
where $R_p$ is the planet radius, $a$ is the semi-major axis. 
The dayside thermal emission can be parameterized as follows 
%
%
\begin{eqnarray}
\label{eq:day_depth}
\delta_{therm} = \left({R_p \over R_s}\right)^2 {F_p(\alpha,\lambda,T_0) \over F_s(\lambda)}
\hspace{2mm},   
\end{eqnarray}
where $R_s$ is the stellar radius, $F_p$ and $F_s$ are the 
wavelength ($\lambda$) dependent fluxes of the planet and the 
star, respectively. In the black body approximation, parameter 
$\alpha$ is used to relate the substellar temperature 
$T_0=T_{eff}\sqrt{R_p/a}$ to the dayside temperature $T_{day}$ 
%
%
\begin{eqnarray}
\label{eq:day_temp}
T_{day}^4 = \alpha T_0^4 \hspace{2mm}, \hspace{2mm} \alpha=(1-A_b)\left({2 \over 3}-\epsilon {5 \over 12}\right)
\hspace{2mm}.   
\end{eqnarray}
The single parameter $\alpha$ comprises the Bond albedo $A_b$ and 
the atmospheric circulation parameter $\epsilon$. For the origin of 
the coefficients in the expression of $\alpha$ we refer to \cite{burrows2008} 
and \cite{cowan2011}. Obviously, separation of $A_b$ and $\epsilon$ is  
not possible by using the occultation alone. However, by measuring 
the phase curve, one may attempt to derive the night side emissivity 
that depends solely on $\epsilon$ \citep{singh2021}. Because of the 
lack of very high quality data required by this method, we opted 
to parameterize the derived geometric albedo depending on the extreme 
limits of $\epsilon$ and omitting the negligible temperature 
decrease due to the expectedly small Bond albedo \citep[e.g.,][]{mallonn2019}. 
Furthermore, as usual, we assumed pure black body radiations both 
for the star and for the planet in the waveband of interest.   
 
%
%
\begin{figure}[h]
\centering
\includegraphics[width=0.40\textwidth]{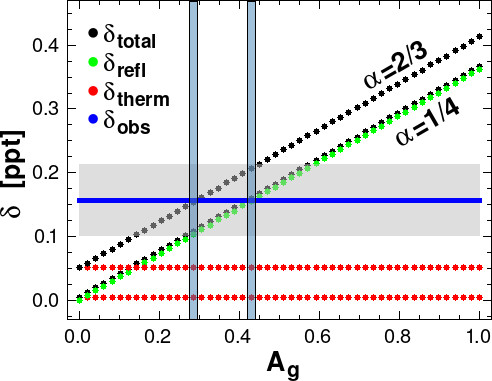}
\caption{Share of the thermal and reflected lights in the total flux 
         change during occultation with varying geometric albedo. 
	 The dots are for two extremes of planetary atmosphere 
	 dynamics with negligible Bond albedo: $\alpha=2/3$ -- complete 
	 lack of circulation, $\alpha=1/4$ -- fully efficient circulation. 
	 The $1\sigma$ error of the observed value is indicated by 
	 the gray-shaded stripe. The vertical stripes show the 
	 resulting geometric albedos.}
\label{Ag}
\end{figure}

The result is shown in Fig.~\ref{Ag}. Although with fully efficient 
circulation ($\alpha=1/4$) the geometric albedo can be as high as 
$A_g=0.43\pm 0.15$, based on the Ks occultation data 
\citep[e.g.,][]{kovacs2019} and several other, more direct studies 
\citep[i.e., those based on full phase curve analyses, such as][]{keating2019} 
it is highly unlikely that WASP-5~b stands out from the other hot 
Jupiters, that mostly have low circulation efficiency. Therefore, 
it is quite reasonable to assume that the true value of $A_g$ is 
closer to the no circulation limit of $A_g=0.29 \pm 0.15$.

%
\section{Conclusions}
\label{sect:summary}
In this work we dealt with the secondary eclipse (occultation) light 
curve of the hot Jupiter WASP-5~b. Our goal was twofold: 
i) derive accurate occultation light curve in the 2MASS Ks band, and, 
ii) use the latest TESS data to obtain the first estimate on the 
occultation depth in the optical. For goal i) we used already published 
Ks photometry by \cite{chen2014} and \cite{zhou2015} and combined these 
with the so far unpublished observations made by the FourStar infrared 
imager of the Baade 6.5~m telescope in 2011. Following the principle of 
``minimum data massage'' we ended up with a high precision occultation 
light curve. The relative flux depression (planet vs star) is 
$2.70\pm 0.14$~ppt, which places WASP-5~b among the top few extrasolar 
planets with Ks occultation light curve of relative precision 
this high. 

We attempted to find the signature of secondary eclipse and phase 
variation in the visible by using the currently available TESS data 
from two visits. Based on the statistical tests presented in 
Appendix~\ref{app_B} we found convincing pieces of evidence that the 
underlying signals are unlikely to have total variation greater than 
$0.20$~ppt. As a result, we accepted the eclipse depth derived from 
the formal fit, i.e., $(0.157\pm 0.056)$~ppt. 

Using these values, our main conclusions on the atmospheric properties of 
WASP-5~b are as follows. 
\begin{itemize} 
\item
Simple black body radiation fails to reach the observed Ks emission 
at the level of $10\sigma$. A similar statement, with somewhat lower 
significance of $4-7$\%, is also true for the band-averaged values of the 
adopted atmospheric models. Detailed atmospheric modeling with strong 
emission features in the Ks band is required to fit the high observed 
emission.  
\item
The value derived for the emission in the TESS waveband shows no 
preference for any of the adopted models with oxygen or carbon enhancements. 
The observed emission value is $\sim 2\sigma$ apart from both models.
\item
From the TESS eclipse depth we found that, depending on the 
circulation model, the geometric albedo $A_g$ is likely in the range of 
$0.29$--$0.43$. This places WASP-5b among the most reflective extrasolar 
planets but with a caveat of the preliminary nature of the detection of 
the secondary eclipse in the optical.  
\end{itemize}

It is also worth mentioning that the ephemeris of the Ks occultation 
light curve further confirms the low (likely zero) eccentricity of the 
orbit, namely $e=0.005\pm 0.015$. Furthermore, the TESS data do not 
suggest the presence of any additional transiting planet larger than 
$\sim 2$ Earth radii with a period between $~30$ and $0.1$ days. 

The Ks waveband is within a relatively easy reach for most of the 
ground-based telescopes with near infrared capabilities. However, 
the time scale and the signal level make extrasolar occultation 
measurement still challenging, due to the combination of the above 
two properties with the local conditions, exhibited via the red noise 
component of the observed photometric time series. Perhaps 
the best way of handling red noise (if there are no other ways to 
filter it out), is to take multiple samplings. The careful combination 
of these samples will reduce both the white and red noise components. 
Due to the sparse sampling from the side of the available data points 
in the different wavebands, measurement accuracy is crucial for 
spectral retrieval. Although this spectral band is (will be)  
available in various space missions 
\citep[JWST now and ARIEL by the end of the decade -- see ][]{tinetti2018}, 
the expected high demand (in particular for JWST) makes ground-based 
observations still very important in supplying high quality data for 
more reliable extrasolar planet atmosphere modeling.

%
%
\begin{acknowledgements}
%
Constructive comments by the referee are appreciated, in particular 
those that have led to a deeper investigation of the significance of the 
secondary eclipse in the TESS band. 
We thank Nikku Madhusudhan for the valuable comments regarding the 
atmospheric modeling of WASP-5b.  
%
This paper includes data collected with the TESS mission, obtained 
from the MAST data archive at the Space Telescope Science Institute 
(STScI). Funding for the TESS mission is provided by the NASA Explorer 
Program. STScI is operated by the Association of Universities for 
Research in Astronomy, Inc., under NASA contract NAS 5-26555.
%
I.D. was supported by the Deutsche Forschungsgemeinschaft 
(DFG, German Research Foundation) -- Project-ID 138713538 -- SFB 881 
(``The Milky Way System'', subproject A03).
%
G.C. acknowledges the support by National Natural Science 
Foundation of China (Grant No. 42075122, 12122308). 
%
Support from the National Research, Development and Innovation 
Office (grants K~129249 and NN~129075) is acknowledged. 
\end{acknowledgements}
%

%
%

%

%
%
\begin{appendix}
\section{Trapezoidal fits to the nightly Ks data}
\label{app_A}
%
%
%
\begin{table*}[h!]
\centering
\begin{minipage}{200mm}
\caption{Trapezoidal fit parameters to the nightly Ks datasets}
\label{fit-per-set}
\scalebox{0.9}{
\begin{tabular}{cccccccc}
\hline
Set & $T_1$   & $T_4$   & $T_{14}$ & $T_{12}$ 
& $\delta$       & $\sigma_{fit}$ & $N$\\
    & (phase) & (phase) & (phase) & (phase) 
& ($\Delta F/F$) & ($\Delta F/F$) & \\
\hline
1   
& $0.47908\pm0.00411$ 
& $0.53234\pm0.00468$
& $0.05326\pm0.00576$
& $0.00766\pm0.00128$
& $0.00293\pm0.00037$
& $0.00356\pm0.00012$
& \phantom{2}$699$\\
2
& $0.46810\pm0.00267$
& $0.53204\pm0.00180$
& $0.06394\pm0.00348$
& $0.00633\pm0.00102$
& $0.00270\pm0.00018$
& $0.00300\pm0.00005$
& $2084$\\
3
& $0.47169\pm0.00310$
& $0.53073\pm0.00444$
& $0.05903\pm0.00426$
& $0.00525\pm0.00084$
& $0.00279\pm0.00022$
& $0.00327\pm0.00007$
& $1732$\\
123
& $0.47146\pm0.00219$
& $0.53102\pm0.00139$
& $0.05956\pm0.00270$
& $0.00528\pm0.00070$
& $0.00270\pm0.00014$
& $0.00317\pm0.00004$
& $4515$\\
\hline
\end{tabular}}
\end{minipage}
\begin{flushleft}
\vspace{-5pt}

{\bf Notes:}
The ingress and egress phases ($T_{1}$, $T_{4}$) can be converted 
into Barycentric Julian Date (TDB standard) by the following formulae: 
$T_{ing,egr}$[BJD]$=T_{cen}+P\times(n+T_{1,4})$, where $n$ is the 
epoch number of the event of interest and $T_{cen}=2458355.50805$ 
is the moment of the transit center, $P=1.6284300$~d 
-- see Sect.~\ref{sect:tess-eph} for more details. The epochs 
are {\em without} correction for orbital light time effect. 
\end{flushleft}
\end{table*}

In this Appendix we give further details of the method of the 
error estimation of the trapezoidal fit. This section serves 
also the closer assessment of the improvement resulted from 
merging the three datasets available in Ks color.  

The individual nightly data by \cite{chen2014}, this paper 
and \cite{zhou2015} are fitted by the trapezoidal model of 
Sect.~\ref{sect:set-2}. The procedure followed for the individual 
datasets is the same as described in Sect.~\ref{sect:merge-123}. 
The starting time series are the light curves that have already 
been filtered out from systematics and major outliers (see 
Fig.~\ref{LC-123}). To include the effect of changing noise 
level (both random and left-in systematics) throughout the night, 
the light curves were binned and the average bin values were used 
as the starting light curve of the simple Monte Carlo simulations. 
These simulations were performed with $500$ noise realizations 
superposed on this ``noiseless'' light curve. The standard deviation 
of the superposed Gaussian noise changed according to the standard 
deviation of the observed data in each bin. 

We used $80$ overlapping bins, i.e., the total time span of a given 
dataset was divided into $40$ equal time segments and the averages 
and standard deviations of the data points belonging to each of these 
segments were calculated. Then, the overlapping sequence started at 
half of the first bin and continued until the half of the last bin. 
The bin width of the second sequence was somewhat shorter, because 
we generated the same number of bins (i.e., $40$) as for the first 
sequence. Finally, all the $80$ bin averages and standard deviations 
were assigned to the $80$ bins of equal length filling in the full 
time span. This method of bin generation was used for all three 
datasets, independently of their lengths. While generating the mock 
light curves for noise estimation, we mapped back the bin statistics 
to the original timebase by choosing the bin values at the given 
time (or phase) value of the unbinned time series. 

The simulated light curves served as inputs for our robust trapezoidal 
fitting routine. The resulting parameters were used to compute their 
standard deviations ($1\sigma$ errors). The trapezoidal parameters 
themselves were obtained from a direct fit to the original (unbinned) 
data. In searching for the best fit, we used simple parameter scanning 
in the following ranges (in the units of the orbital phase):
$ 0.45151 < T_{ing} < 0.49151 $, 
$ 0.04747 < T_{14}  < 0.07120 $, 
$ 0.00409 < T_{12}  < 0.00613 $, 
corresponding to $\pm 0.02$ phase change for $T_{ing}$ with respect 
to a finally accepted best-fit value. The scanning ranges in 
$T_{14}$ and $T_{12}$ correspond to $\pm 20$~\% relative variations 
(again, with respect to the same final best-fit values). Except for 
$T_{12}$, we had very rare hits at the limiting values. The more 
frequent hits for $T_{12}$ can be tolerated on the ground of the 
limits imposed by the known system parameters.  
 
The result is shown in Table~\ref{fit-per-set}. To make the comparison 
easy, we copied the result of the analysis of the merged data from 
Table~\ref{merged_par} in the fourth row. In the last but one column 
we show the standard deviation of the fit to the original data. 
The errors of the standard deviation of $\sigma_{fit}$ come from the 
Monte Carlo simulations and indicate the data point number dependence 
of the statistical stability of the realizations. 

We see that the merged data yield a considerable improvement with 
respect to all datasets. With an overall error decrease of $\sim 20$\%, 
a relative modest improvement can be observed for set-2. For the other 
sets we find improvement in the range of $\sim 50$--$300$\%. It is 
also observable that while our transit depth error is in agreement 
with the one derived by \cite{zhou2015} there is nearly a factor of 
two difference between our error of $0.037$~ppt, vs that of \cite{chen2014}. 
It is quite likely, that the latter estimate ($0.062$~ppt) is closer 
to the real error, since their estimate follows the full sequence of 
light curve evaluation (which is certainly very sensitive to the method 
of systematics treatment), whereas our estimate relies on their 
``final product''. However, at the end, from the point of view of the 
merged data analysis, the weight of this dataset is relatively small, 
due to the small number of data points. 
   
%
%
%
\section{Testing occultation and phase variation limits from TESS}
\label{app_B}
We performed eclipse and phase variation analyses on the 
full set of TESS light curves -- including sectors $02$ and $29$. 
With the option of polynomial filtering (see Sect.~\ref{sect:tess}), 
we have four types of datasets to be analyzed (SAP and PDC, with 
or without polynomial filtering). 
The time series models were incomplete -- i.e., they contained 
either an occulation or a phase variation signal. This is a 
reasonable approximation in the high noise regime, when focusing 
on the detection limits of signals with very different time 
dependences.  

\begin{figure}[h!]
\centering
\includegraphics[width=0.45\textwidth]{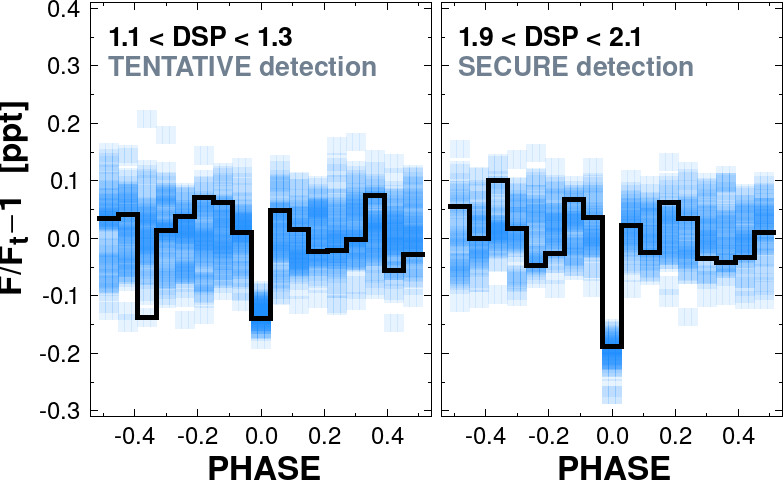}
\caption{Parametrization of the phase-folded, binned test signals 
         by DSP (see Eq.~\ref{eq:dsp}). Several realizations of the 
	 signal described in the text and in the caption of Fig.~\ref{box_det} 
	 were used with injected box depth of $\delta=0.15$~ppt. The 
	 width of the bins is equal to the eclipse duration of WASP-5, 
	 and are distributed according to the eclipse phase (centered 
	 arbitrarily at phase zero). In both cases one realization is 
	 shown by black line to follow bin-by-bin fluctuations more 
	 easily. To avoid unnecessary jamming, we show only realizations 
	 with the largest negative bin average at the eclipse center. 
	 The difference in the overall eclipse depth in the two cases 
	 is due to the different DSP cutoffs used.}
\label{dsp_vs_lc}
\end{figure}

For the eclipse dip search, the folded light curve was  
approximated by a bin model, with equal widths for all bins, 
equal to that of the transit. The distribution of the bins was 
fixed, as given by the predicted phase of the secondary eclipse. 
The bin average yielding the largest flux depression and the 
center phase of that bin yields, respectively, the estimated 
eclipse depth $\delta$ and eclipse phase $\psi$. To characterize 
the quality of this representation of the eclipse, we used the 
dip significance parameter (DSP) as defined in \cite{kovacs2020} 
%
%
\begin{eqnarray}
\label{eq:dsp}
{\rm DSP} = {\delta \over \sqrt{e_{\delta}^2 + \sigma_{ab}^2 + \sigma_{db}^2}} 
\hspace{2mm},   
\end{eqnarray}
where $e_{\delta}$ is the error of the eclipse depth $\delta$, 
$\sigma_{ab}$ is the standard deviation of the bin averages for 
the bins outside the eclipse, $\sigma_{db}$ is the standard 
deviation of the successive differences between the same bin 
averages. The last two terms are devoted to account for both 
regular, smooth variations ($\sigma_{ab}$) and irregularities 
originating from pure noise at the phase scale of the bins 
($\sigma_{db}$). The DSP value associated with the deepest 
bin average (i.e., the largest $\delta$) was used to characterize 
the quality of the eclipse solution for a given dataset. 
It is noted that for a complete signal, DSP is limited 
by the second term in the denominator, due to the near equality 
of the eclipse depth and the total amplitude of the phase 
variation. Figure~\ref{dsp_vs_lc} shows the relation between the 
light curve quality and the above parametrization. 

%
%

The phase variation search was performed by a simple two-parameter 
($A$ and $\phi_0$) least squares fit of a cosine function 
$A\cos(2\pi(\phi-\phi_0))$. With a good approximation, the 
amplitude $A$ of the cosine is half of the occultation depth 
\citep[e.g.,][]{daylan2021}. The quality of the cosine model 
is characterized by the signal-to-noise ratio 
${\rm SNR}=A/\sigma_{fit}$, which -- as expected -- for low-SNR 
signals, like the ones studied here, is tightly correlated with 
$A$.    

The results of the above separate fits for the four types of 
datasets are displayed in Table~\ref{occ_det4}. Although there 
are differences among the datasets, the properties of the 
signals are largely stable. The box fit yields remarkable 
coincidence with the expected occultation center. On the 
other hand, the maximum of the cosine is systematically shifted, 
and precedes the occultation center in phase by $\sim 0.36$. 
The amplitude of the phase variation and the eclipse depth 
are consistent within the error limit: 
$\langle \delta \rangle = 0.157\pm 0.056$~ppt vs 
$\langle 2A \rangle = 0.113\pm 0.041$~ppt.  

%
%
\begin{table*}[h!]
\centering
\caption{Best-fit bin and cosine signals in the TESS data}
\label{occ_det4}
\scalebox{1.0}{
\begin{tabular}{ccc|ccc|cccc}
\hline
\multicolumn{3}{c}{DATA} & \multicolumn{3}{c}{BIN fit} & \multicolumn{3}{c}{COSINE fit} \\
\hline
Type  &  Pol  & $\sigma$  [ppt]&  $\Delta\varphi$  &  $\delta$ [ppt] & DSP & $\Delta\varphi$  & $A$ [ppt] & SNR \\
\hline
\hline
 SAP &  0 & 2.84 & 0.00 & $0.155 \pm 0.062$ & 1.09 & $-0.36$ & $0.043 \pm 0.023$ & 1.90 \\
 SAP &  1 & 2.49 & 0.00 & $0.164 \pm 0.054$ & 1.11 & $-0.37$ & $0.063 \pm 0.021$ & 3.07 \\
 PDC &  0 & 2.35 & 0.00 & $0.161 \pm 0.054$ & 1.24 & $-0.38$ & $0.066 \pm 0.019$ & 3.50 \\
 PDC &  1 & 2.32 & 0.00 & $0.149 \pm 0.053$ & 1.24 & $-0.32$ & $0.053 \pm 0.018$ & 2.94 \\
\hline
\end{tabular}}
\begin{flushleft}
\vspace{-5pt}
{\bf Notes:}
Pol: $1/0$ for with or without polynomial filtering; 
$\sigma$: standard deviation of the input time series; 
$\Delta\varphi=\psi-\varphi_0$ for the bin model and  $\Delta\varphi=\phi_0-\varphi_0$ 
for the cosine model; 
$\delta$: eclipse depth;
DSP: dip significance parameter;
A: amplitude of the cosine fitted; 
SNR: signal-to-noise ratio for the cosine fit. See text for additional 
information on the symbols.  
\end{flushleft}
\end{table*}

It is important to address the statistical significance of these, 
obviously near noise-level detections. To do so, we performed simple 
Monte Carlo simulations by generating the following type of signals 
%
%
\begin{eqnarray}
\label{eq:signal}
x(i) = (1.0 + G(i))\times(1.0 + S(i)) \ ; \quad i=1,2, ..., n  
\hspace{2mm},   
\end{eqnarray}
where $G(i)$ is an uncorrelated Gaussian noise  with standard 
deviations shown in Table~\ref{occ_det4}. Function $S(i)$ is 
either a simple box or a cosine function, representing the 
secondary eclipse or the phase variation. The box model is 
zero out of the eclipse and $-\delta$ within the eclipse, 
with the box centered at the phase of the assumed eclipse, 
$\varphi_0$ ($0.5$ phase after the transit). Probability 
density functions (PDFs) and cumulative distribution functions 
(CDFs) were computed for the phase differences 
(i.e., $\Delta\varphi=\psi-\varphi_0$), DSP and A values.   

%
%
\begin{figure}[h!]
\centering
\includegraphics[width=0.40\textwidth]{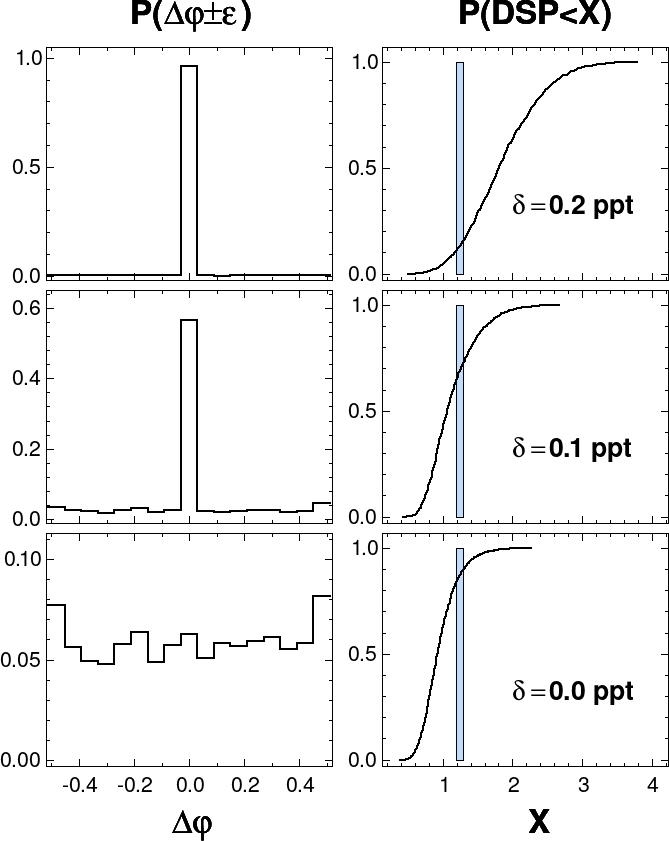}
\caption{Injected box test of pure Gaussian noise with $\sigma=2.32$~ppt 
         and $n=33256$ data points. 
         {\em Right column:} CDF of DSP for various injected box 
	 depths $\delta$. Vertical bars show the observed value 
	 ($6^{th}$ column of Table~\ref{occ_det4} in row $4$).
         {\em Left column:} PDF of the phase difference between the 
	 test box center and the calculated secondary eclipse center 
	 from the bin model for the same box depths as shown in the 
	 corresponding panel on the right. The bin width is the total 
	 transit duration, i.e., in phase units $\epsilon=0.030$.}
\label{box_det}
\end{figure}
%

%
%
\begin{figure}[h!]
\centering
\includegraphics[width=0.40\textwidth]{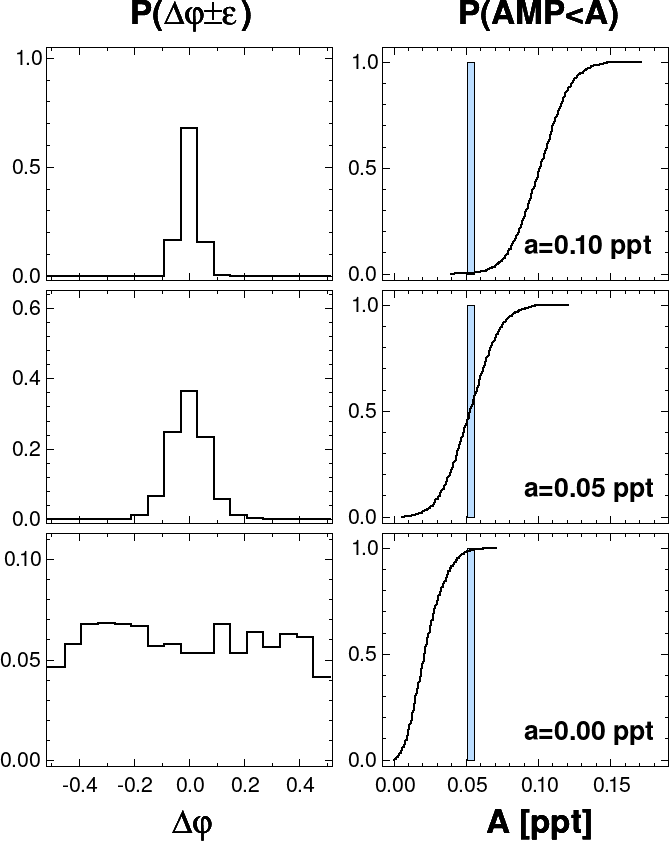}
\caption{As in Fig.~\ref{box_det} but for the injected cosine test. 
         The amplitudes $a$ of the injected cosine functions were 
	 chosen to be compatible with the corresponding box test 
	 (i.e., $2a=\delta$).}
\label{sin_det}
\end{figure}

The primary goal of this investigation is to pose an upper limit 
on the underlying signal by using the near noise-level values 
derived above. Therefore, we computed two basic quantities. 
In the presence of injected signals of varying amplitudes 
and box depths we computed:  
a) the occurrence rates of box depths and cosine amplitudes 
less than the observed values; 
b) the occurrence rates of the dip center and cosine maximum 
phases within the proximity of the predicted secondary eclipse. 

As an example, for three different injected box signals, 
Fig.~\ref{box_det} shows the PDF of the phase differences and 
the CDFs of the DSP values. The phases seem to converge quite 
quickly (i.e., even for boxes as shallow as $0.1$~ppt, more 
than half of the cases hit the near proximity of the expected 
phase). It is interesting to note that there is surplus of 
occurrences at the edges in the phase distribution. It is 
especially visible in the pure noise ($\delta=0.0$) simulations. 
The reason of this excess is the smaller number of data points 
in the bins at the end-phases of the folded time series (the bin 
occupancy depends on the positioning of the occultation and the 
width of the eclipse). This leads to more fluctuating values 
at the edges, and therefore, higher chance of being selected 
as the ``best-fit'' for the box model. 

The DSP values are less sensitive to a small underlying signal. 
They are still near the pure noise values and they become more 
distinctive from these only if the box depths become deeper than 
$0.15-0.20$~ppt. 

A similar test performed on the same type of dataset with injected 
cosine signals indicate the opposite effect (see Fig.~\ref{sin_det}). 
The phase settles at a far lower pace but the amplitude of the 
signal becomes more quickly detectable. For example, if the underlying 
cosine signal had an amplitude of $0.1$~ppt (corresponding to 
an eclipse depth of $0.2$~ppt), then the probability that we 
can detect a component as low as observed is less than $0.003$. 

\begin{figure}[h!]
\centering
\includegraphics[width=0.40\textwidth]{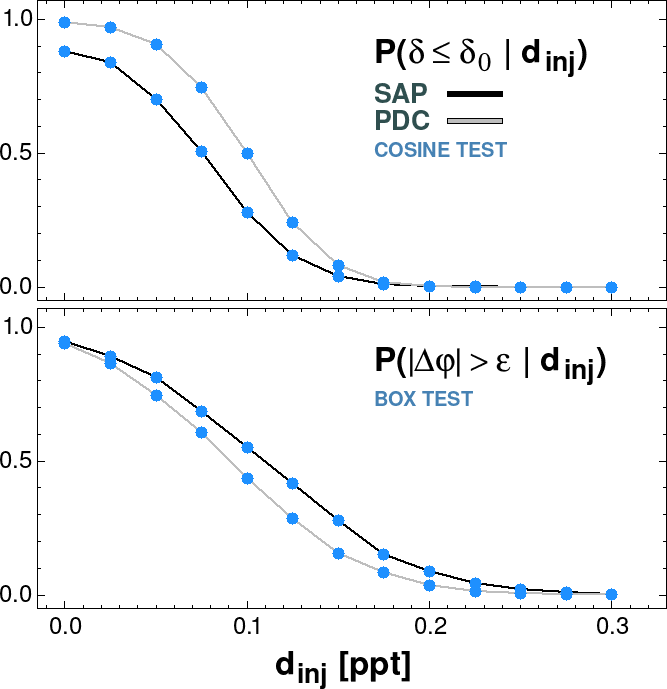}
\caption{{\em Upper panel:} Occurrence rate of the detection of 
         cosine amplitudes $\delta/2$ less than $\delta_0/2$ in 
	 the presence of injected cosine amplitudes 
	 d$_{\rm inj}/2$. The Gaussian components of the mock 
	 signals were generated by using the standard deviations 
	 of the SAP and PDC time series (first and last rows in 
	 Table~\ref{occ_det4}). The $\delta_0$ values refer to 
	 the respective total amplitudes in the same table (i.e., 
	 $0.086, 0.106$~ppt for the SAP and PDC data, respectively).   
         {\em Lower panel:} Occurrence rate of phase difference  
	 $\Delta\varphi$ for the same noise models as 
	 above but injected by box signals of depths d$_{\rm inj}$ 
	 and widths of $2\epsilon=0.060$.} 
\label{det_p}
\end{figure}

We can assess the likelihood of the various underlying signals 
in the currently available TESS data by using the data settings 
(i.e., data point distribution, N and $\sigma$) for the two extreme 
data types shown in Table~\ref{occ_det4}. These settings 
correspond to the PDC and SAP fluxes, respectively, with and 
without polynomial detrending. We utilize the detection power 
of the phase of the bin search and the amplitude sensitivity 
of the Fourier fit. The upper panel of Fig.~\ref{det_p} shows the 
occurrence rate of the observed total amplitude as a function 
of the underlying (injected) signal amplitude. The lower 
panel exhibits the likelihood that the best-fitting bin phase is 
not being in the close proximity of the predicted phase. 

%
%

These plots suggest that the underlying phase variation should 
have a total amplitude less than $0.20$~ppt with a probability 
of more than $99.7$\%, because otherwise it would have a 
probability less than $0.3$\% to get a cosine amplitude as small 
as given in Table~\ref{occ_det4}.  
Although at a somewhat lower level of significance, this result 
is corroborated by the frequency of the correct phase hits in 
the box test. Depending somewhat on the data type, the probability 
of not hitting the correct phase for an underlying boxy eclipse 
signal of depth $0.20$~ppt is between $4$\% and $9$\%. We note 
that the average of the eclipse depths derived from the observed 
data and the associated $1\sigma$ formal error yield an upper 
limit of $0.157+0.056=0.213$~ppt, close to the high-significance 
limit obtained above. This, together with the stability of 
the observed eclipse phases, lend further support to the 
tentative detection of the secondary eclipse for WASP-5b from 
the TESS data.   

\end{appendix}

\end{document}